\documentstyle[aaspp]{article}
\def\kms {km~s$^{-1}$}

\newcommand \nnhp{\mbox{N$_2$H$^+$}}
\newcommand \hcop{\mbox{HCO$^+$}}
\begin{document}
\title{How to Identify Pre-Protostellar Cores}
\author{Erik M. Gregersen\altaffilmark{1}}
\affil{Department of Physics and Astronomy, McMaster
University, Hamilton, ON L8S 4M1, Canada}
\author{Neal J. Evans II\altaffilmark{2}}
\affil{Department of Astronomy, The University of Texas at Austin,
       Austin, TX 78712--1083}
\altaffiltext{1}{Electronic mail: gregerse@physun.physics.mcmaster.ca}
\altaffiltext{2}{Electronic mail: nje@astro.as.utexas.edu}
\centerline{\footnotesize {\LaTeX}ed at \number\time\ min., \today}

\begin{abstract}

We have observed the HCO$^{+}$ $J=3-2$ line toward 17 starless cores selected 
from the list of Ward-Thompson et al.\ (1994). Six of these cores have line 
asymmetries indicative of collapse.  The excess of blue-skewed profiles over 
red-skewed profiles is at least as large as that found in
samples of Class 0 and early Class 
I sources.  The observed line profiles have the same narrow linewidths and 
small peak temperatures predicted for young sources in evolutionary models, but 
the blue/red ratios, like those of older sources, are higher than models 
predict.  The infall signature also occurs over large scales, suggesting that 
these cores have overall inward motions.  We have divided these starless cores 
into two groups based on the continuum photometry of Ward-Thompson et al.\ and 
our HCO$^{+}$ data.  We find stronger HCO$^{+}$ emission among the cores 
detected in the submillimeter and all the blue-skewed line profiles are in this 
group, supporting the suggestion of Ward-Thompson et al.\ that these are the 
pre-protostellar cores.
\end{abstract}


\section {INTRODUCTION}

For years, the protostellar collapse stage had been an impenetrable mystery,
until Walker et al.\ (1986) observed line profiles indicative of collapse in 
IRAS 16293-2422.  Menten et al.\ (1987) disputed this interpretation and 
claimed that the asymmetric line profiles of Walker et al.\ were caused by 
rotation, but Zhou (1995) later modeled IRAS 16293-2422 as collapse with 
rotation. Zhou et al.\ (1993) observed B335, a slowly rotating source, and  
modeled its blue-peaked profiles as inside-out collapse (Shu 1977).  Andr\'e et 
al.\ (1993) extended the tripartite taxonomy of young stellar objects to 
include Class 0 objects (very embedded sources, such as B335 and IRAS 16293). 
Andr\'e and Montmerle (1994) found that the Class 0 sources were more embedded 
than Class I sources and inferred that they had not yet accreted most of their 
mass.  Spectral line surveys of Class 0 sources (Gregersen et al.\ 1997, 
Mardones et al.\ 1997), found nearly a third to half of Class 0 objects 
displayed asymmetries in optically thick lines like those seen in B335 and IRAS 
16293-2422.

However, the earliest phase of the collapse process, the transition between 
the quasi-static core formation and the beginning of infall onto a central
object, is poorly understood. Beichman et al.\ (1986) examined the IRAS data 
for 95 cloud cores previously surveyed by Myers et al.\ (1983), Myers and 
Benson (1983) and Benson (1983) in $^{13}$CO, C$^{18}$O and NH$_{3}$ and found 
that half had IRAS sources, which they deduced as arising from protostars.  
Ward-Thompson et al.\ (1994) observed 17 cores from Beichman et al.\ that have 
no IRAS sources.  They detected 12 of these cores in the submillimeter and used 
maps to study the density profiles of 5 cores.  Since these objects lacked IRAS 
sources, it is believed that protostars have not yet formed.  From statistical 
arguments about the lifetime of these cores and the fact that the observed 
density profiles are similar to those predicted by ambipolar diffusion models, 
Ward-Thompson et al.\ identified these starless cores as in the ambipolar 
diffusion phase and pre-protostellar.  This stage precedes the Class 0 phase
and is sometimes referred to as the pre-protostellar core stage.

We observed the objects surveyed by Ward-Thompson et al.\ 
using the HCO$^{+}$ $J=3-2$ line, a line that readily 
displays an asymmetry indicative of protostellar collapse to see if an early 
collapse phase could be found. Lee et al. (1999) have completed a similar 
survey using CS and \nnhp\ lines.

\section {OBSERVATIONS AND RESULTS}

We observed the 17 starless cores listed in Table 1 in the HCO$^{+}$ $J=3-2$ line with the 
10.4-m telescope of the Caltech Submillimeter Observatory (CSO)\footnote{The 
CSO is operated by the California Institute of Technology under funding from 
the National Science Foundation, contract AST 90--15755.} at Mauna Kea, Hawaii 
in March 1995, December 1995, June 1996, July 1998, December 1998 and July 
1999.  
We used an SIS receiver (Kooi et al.\ 1992) with 
an acousto-optic spectrometer with 1024 channels and a bandwidth of 49.5 MHz as the backend.  The frequency resolution ranged from slightly less than 3 
channels, 0.15 \kms\ at 267 GHz, for the 1995 observations to closer to 2 
channels, 0.12 \kms\ at 267 GHz, 
for the 1998 observations.  The antenna temperature, $T_{A}^{*}$, was obtained from 
chopper-wheel calibration.  
Information about the observed lines is listed in Table 2. 
Planets were used as calibration sources for calculating the main 
beam efficiency.  Data from separate runs were resampled to the resolution of 
the run with the worst frequency resolution before averaging.  A linear 
baseline was removed before scans were averaged.  

Line properties are listed in Table 3.  For lines without two clearly
distinguished peaks, $T_{A}^{*}$, the peak temperature, $V_{LSR}$, the line 
centroid, and $\Delta$V, the line width, were found by fitting a single 
Gaussian to the line profile.  For lines with two clearly distinguished peaks, 
we list 
two values of $T_A^*$ and $V_{LSR}$, one for each peak, and we give one value 
for the line width, which is the full width across the spectrum at the 
temperature where the weaker peak falls to half power.

We observed 17 sources in this survey.  All of the sources were observed in
the HCO$^{+}$ $J=3-2$ line.  Six sources were also observed in the 
H$^{13}$CO$^{+}$ $J=3-2$ line.  Six sources showed a blue asymmetry in the 
HCO$^{+}$ $J=3-2$ line (Figure 1).  Eight sources showed symmetric lines 
(Figure 2) and three sources were not detected.  The spectra in Figures 1 and 2 
are from the central position except that of L1689B which is from 
($-$15\arcsec, 15\arcsec), which we chose because it was the strongest position.

\section {Individual Sources}

\subsection {L1498}

Wang (1994) observed absorption in the H$_{2}$CO 6 cm line against the 
cosmic microwave background radiation similar to that
observed in B335.  Kuiper et al.\ (1996) posited that this core is quasi-static or
slowly contracting and that the outer envelope is growing.  They also
concluded that this core could collapse within the next 5 $\times$ 10$^{6}$
years.  Wolkovitch et al.\ (1997) determined that this core was extremely 
quiescent based on its narrow CCS line widths.
The HCO$^{+}$ $J=3-2$ line (Figure 2) shows no asymmetry and is at the same velocity
as the N$_{2}$H$^{+}$ and C$_{3}$H$_{2}$ lines observed by Benson et al.\
(1998).

\subsection {L1506}

The HCO$^{+}$ $J=3-2$ spectrum (Figure 2) shows one component at 7.5 \kms\ with
a possible second component at 9 \kms.

\subsection {L1517C}

The HCO$^{+}$ $J=3-2$ line (Figure 2) is too weak to detect an asymmetry.

\subsection {L1517A}

We observe a slight blue shoulder in HCO$^{+}$ $J=3-2$ (Figure 2).

\subsection {L1512}

Caselli et al.\ (1995) observed the hyperfine components of the N$_{2}$H$^{+}$
$J=1-0$ line and found that a single excitation temperature could not fit the
spectra, an anomaly usually seen in starless cores.  The HCO$^{+}$ $J=3-2$ 
line (Figure 2) is symmetric and is at the same velocity as the 
N$_{2}$H$^{+}$ and C$_{3}$H$_{2}$ lines (Benson et al.\ 1998).

\subsection {L1544}

Myers et al.\ (1996) modeled the H$_{2}$CO $J=2_{12}-1_{11}$ line as arising
from infall.  Tafalla et al.\ (1998) found that their CS $J=2-1$ observations 
could be modeled as arising from inward motions, but those inward motions are
not consistent with the predictions of the Shu (1977) inside-out collapse 
model.  Williams et al.\ (1999) observed similar infall speeds in 
N$_{2}$H$^{+}$ $J=1-0$ on scales of 10\arcsec\ to those observed by Tafalla et
al.  Ohashi et al.\ (1999) mapped this core in CCS $J_{N}=3_{2}-2_{1}$ and 
observed both infall and rotational motion.   Ciolek and Basu (2000) have 
modeled the observations of Tafalla et al.\ and Williams et al.\ in the
context of ambipolar diffusion.  The HCO$^{+}$ $J=3-2$ spectrum is 
blue-peaked with the H$^{13}$CO$^{+}$ $J=3-2$ line peaking in the dip (Figure 
1).  The N$_{2}$H$^{+}$ peaks between the H$^{13}$CO$^{+}$ line and the 
HCO$^{+}$ peak while the C$_{3}$H$_{2}$ peaks on the H$^{13}$CO$^{+}$ line 
(Benson et al.).

\subsection {L1582A}

The HCO$^{+}$ $J=3-2$ line is symmetric (Figure 2).  The N$_{2}$H$^{+}$ 
line of Benson et al.\ has the same peak velocity as the HCO$^{+}$ line. 
There is a hint of a blue shift relative to the \nnhp\ velocity, but not
enough to warrant inclusion as an infall candidate.

\subsection {L134A}

The HCO$^{+}$ $J=3-2$ spectra displays a symmetric line (Figure 2).  
The N$_{2}$H$^{+}$ and C$_{3}$H$_{2}$ lines (Benson et al.) peak 
at the blue edge of our HCO$^{+}$ line. Because those lines were observed
115\arcsec\ from our position, we disregard this source in the statistical
discussion in \S 4.2.

\subsection {L183}

Fulkerson and Clark (1984) used the H$_{2}$CO 6 cm line to model the 
density distribution as an inverse square law.  The CS $J=2-1$ line 
observed by Snell et al.\ (1982) is similar in velocity and shape to our 
HCO$^{+}$ $J=3-2$ spectra (Figure 1).  The HCO$^{+}$ $J=3-2$ line is 
self-absorbed with the blue peak slightly stronger.  The H$^{13}$CO$^{+}$, 
N$_{2}$H$^{+}$ and C$_{3}$H$_{2}$ lines (Benson et al.) peak in the 
self-absorption dip.   

\subsection {L1696A}

The HCO$^{+}$ $J=3-2$ line is symmetric with a faint blue shoulder 
(Figure 2).  The peak velocity corresponds to that of the optically
thin lines of Benson et al.

\subsection {L1689A}

We observe a broad, blue-skewed line in HCO$^{+}$ $J=3-2$ (Figure 1).

\subsection {L1689B}

The spectra for the HCO$^{+}$ $J=3-2$ line is blue skewed with the
H$^{13}$CO$^{+}$ $J=3-2$ line peaking to the red of the peak velocity 
of the HCO$^{+}$ line (Figure 1).

\subsection {L63}

The HCO$^{+}$ $J=3-2$ line is strongly blue-skewed and the H$^{13}$CO$^{+}$
line peaks in the middle of the HCO$^{+}$ line (Figure 1).  The 
N$_{2}$H$^{+}$ and C$_{3}$H$_{2}$ lines (Benson et al.) are at 
the velocity of the red edge of the blue peak.

\subsection {B133}

Hong et al.\ (1991) have mapped this core in $^{12}$CO and $^{13}$CO $J=1-0$.  
The HCO$^{+}$ $J=3-2$ line is slightly blue-skewed (Figure 1).  The 
N$_{2}$H$^{+}$ and C$_{3}$H$_{2}$ lines of Benson et al.\ peak at the red edge 
of the blue peak.

\subsection {Nondetections}

The HCO$^{+}$ $J=3-2$ line was not detected down to a limit of 0.05 K in
L1495D, L1521A and L1517D.

\section {ANALYSIS}

\subsection {New collapse candidates in \hcop}
 
Six of these cores, L1544, L1689A, L1689B, L183, L63 and B133, have line 
profiles with blue asymmetry.  A blue asymmetry in an optically thick line like 
HCO$^{+}$ $J=3-2$ can be caused by protostellar collapse (Leung and Brown 1977, 
Zhou \& Evans 1994).  We observed H$^{13}$CO$^{+}$ $J=3-2$ in all of these 
sources to find the rest velocity and we also used the N$_{2}$H$^{+}$ and 
C$_{3}$H$_{2}$ observations of Benson et al.\ to provide optically thin line 
velocities.  If optically thin lines have the same velocity as the dip of the 
double-peaked line, the dip is self-absorption from the ambient cloud and not 
caused by two separate velocity components blended together.  

For such narrow lines as seen in these sources, 
the rest frequencies of lines become an issue.
The frequency of the HCO$^{+}$ $J=3-2$ line is known to within an uncertainty
of 0.01 MHz, but the uncertainty of the H$^{13}$CO$^{+}$ $J=3-2$ is potentially
large. The standard value in the JPL data base (Pickett et al. 1998,
http://spec.jpl.nasa.gov) is 260.255478 GHz.
It is calculated from measurements of the two lower lines (Woods
et al. 1976, Bogey et al. 1981), making it impossible to estimate
uncertainties in the $J=3-2$ line frequency.
Comparing the velocities
of sources done in H$^{13}$CO$^{+}$ here and elsewhere (Gregersen et al.\ 1997, 2000) with line velocities of the same sources in N$_{2}$H$^{+}$,
on average the H$^{13}$CO$^{+}$ line is $0.16\pm0.04$ \kms\ 
to the red of the
N$_{2}$H$^{+}$ line.  Since it is unlikely that the H$^{13}$CO$^{+}$ would
be consistently red-shifted from the N$_{2}$H$^{+}$, we have shifted our 
H$^{13}$CO$^{+}$ spectra by 0.16 \kms\ to the blue and propose a
frequency of 260.255617$\pm$0.000035 GHz for the H$^{13}$CO$^{+}$ $J=3-2$ line,
0.139 MHz higher than the standard value.

For four of
these sources, L1544, B133, L183 and L1689B, the H$^{13}$CO$^{+}$ line does peak in
the self-absorption dip, one of the conditions that must be met before a core
can be called a candidate for protostellar collapse.  In L63, the H$^{13}$CO$^{+}$
coincides with the dip but the N$_{2}$H$^{+}$ lies between the dip and the
red peak. 
We did not
detect H$^{13}$CO$^{+}$ in L1689A,
so we cannot make a claim about its quality as a collapse candidate.

How do the conclusions from \hcop\ compare to those of Lee et al. (1999)
who observed CS and \nnhp ? They did not observe B133 or L1689A, but
they found L1544, L1689B and L183 (their position for L183B is closest to the 
core we call L183) to be infall candidates.  There are slight differences
between the CS and \hcop\ spectra.  L1689B has a blue peak with a red shoulder
in \hcop\ and double-peaked profile with a strong blue peak in CS.  L1544
has a strong blue peak in \hcop\ but two peaks of equal strength in CS.   L183 
has similarly shaped spectra in both lines.  Lee et al. (1999) observed
a red peaked profile 
in L63, but their position differs substantially from ours.
In general, the different tracers agree reasonably well.

\subsection {Line profile statistics}

Optically thick lines in collapsing sources show a double-peaked line profile 
with the blue peak stronger than the red peak.  There are two ways of 
quantifying the asymmetry of the line.  One could use the ratio of the 
strengths of the two peaks or the asymmetry parameter (Mardones et al.\ 1997),
$\delta V = (V_{thick} - V_{thin}) \ \Delta v_{thin}$, where $V_{thick}$ is the 
velocity of the peak of the optically thick line, $V_{thin}$ is that of the 
optically thin line and $\Delta v_{thin}$ is the linewidth of the optically 
thin line. We list the results in Table 4.  We do not list the results for all 
the objects where HCO$^{+}$ was observed but only those objects for which we 
were able to establish a rest velocity from H$^{13}$CO$^{+}$, N$_{2}$H$^{+}$ or 
C$_{3}$H$_{2}$.   L134A is listed in parentheses because its \nnhp\ spectrum
was observed 115\arcsec\ from our \hcop\ spectrum.  All the other sources for 
which we used the \nnhp\ line to determine the asymmetry were observed within
40\arcsec\ of our \hcop\ position.
In Table 5, we list the 
number of blue, symmetric, and red sources as determined by the asymmetry 
parameter and visual inspection of the line profiles.  Following the 
classification of Mardones et al.\, sources with asymmetry $<$ $-0.25$ are 
blue, while those with asymmetry $>$ 0.25 are red.  The excess, which 
characterizes how blue is a sample of objects, is the number of blue sources 
minus the number of red sources divided by the total number of sources.  The 
excess as determined from the asymmetry parameter leaving aside L134A is 0.66, 
higher than those derived for samples of Class 0 and I sources, 0.25 and 0.39 
respectively, using the \hcop\ $J=3-2$ line (Gregersen et al.\ 2000).  
Because the number of sources in our sample is small, it is unclear whether
the larger excess is significant, but it appears to be at least as large
as in the later stages. 
Lee et 
al.\ found the mean $\delta V$ of starless cores to be $-0.24\pm$0.04 not so 
different from the mean $\delta V$ for Class 0 sources, $-0.28\pm$0.10 
(Mardones et al.).  For the HCO$^{+}$ observations of all three categories,
the mean $\delta V$ using the H$^{13}$CO$^{+}$ $J=3-2$ line frequency proposed
in the previous section 
are $-0.34\pm$0.13, $-0.11\pm$0.09 and $-0.17\pm$0.08 for
starless cores, Class 0 and Class I objects, respectively.  So on the basis of 
line asymmetries, these starless cores cannot be distinguished from an older, 
protostellar population.  

\subsection{Are these young sources?}

Zhou (1992) modeled CS lines in cores that have not yet formed a protostar but 
are evolving toward the singular isothermal sphere, the starting point of the 
Shu (1977) collapse model.  He found CS lines with no asymmetry, narrow line 
widths that increased after collapse began and peak line temperatures that 
increase until collapse begins.  Gregersen et al.\ (1997) did evolutionary 
models of HCO$^{+}$ $J=3-2$ in collapsing clouds and found that such line
parameters as peak temperature, line width, and blue-red asymmetry increase
with time to a maximum value and then decline.  Therefore, if we observed 
sources that were just beginning to collapse, these sources would have 
HCO$^{+}$ $J=3-2$ lines that have slight asymmetry, small peak temperatures and
narrow line widths.  If we extend these results backward to t = 0, to the 
beginning of collapse, we should expect similar line parameters to those of 
Zhou (1992).  Also, since the luminosity of the protostar does not rise 
immediately, there is some time after collapse begins but before a source would 
have been detected by IRAS, so the early stages of the evolutionary models of 
Gregersen et al.\ could be applied to ``pre-protostellar" objects.  In these 
models, important processes like cloud chemistry and protostellar heating that 
increase the line width and peak temperatures are glossed over or 
simplified, so the resulting evolutionary ``tracks" are best seen as rough 
sketches of line parameter evolution. In Figure 3, we plot peak temperature and 
linewidth versus the blue-red ratio for the two abundance distributions modeled 
in Gregersen et al.  The evolutionary ``tracks" in both plots go roughly from 
the lower left corner to the upper right.  We see that the starless 
cores are congregated toward the lower (i.e. ``younger") half of the plot and 
the Class 0 and I sources which have IRAS sources extend further to the upper 
ends of the two plots.  

However, there are indications that some of these sources have already begun
to collapse in a different way than the Shu model predicts.  Several of the 
starless cores have blue/red ratios that are too large for early 
collapse and more like the extreme blue-red ratios seen in more evolved 
Class 
0 sources.  In fact, B133 displays the most extreme ratio between the blue and 
red peak of any source in any class.  The blue-red ratios for this and other 
sources are too high to arise in a Shu model.  Mardones (1998) modeled several
alternative velocity fields and found that a model in which the entire cloud 
collapses produces the highest blue-red ratio.  These large blue-red ratios are 
further discussed in the next section.

\subsection {Extended infall signatures}

We have partially mapped five cores that showed asymmetric line profiles to 
observe the extent of the asymmetry.  The blue-skewed profiles typically 
stretch over a large area for such young cores.  For example, in L1689B (Figure 
4), the self-absorption or blue-skewed lines stretches over roughly 0.04 by 
0.03 parsecs. There are some red profiles in the southeast area of the map.
For L63 (Figure 5), the blue skewness is seen in an area 0.07 by 0.03 
parsecs. We show these two cores because these are the largest maps we have 
done with good signal-to-noise. Tafalla et al.\ (1998) have commented on this 
large extent of the infall signature in L1544 (map in Figure 6).  They showed 
that if L1544 was undergoing inside-out collapse, it would have produced a 
protostar easily visible to IRAS and that such an early, large-scale infall 
does not fit with most theories of protostellar collapse.  

These observations of extended infall signatures in \hcop\ combined with the 
extreme blue-red ratios mentioned in the last section strengthen the case for 
extended inward motions.  Myers and Lazarian (1998) have proposed that 
turbulent dissipation in a cloud produces an overall implosion starting from 
the core exterior as an explanation for extended infall signatures.  
Alternatively, Li (1999) has found evidence for extended inward motions
at speeds up to half the sound speed in calculations of core formation 
in weakly magnetized clouds.  Similar speeds also appear in the model of
Ciolek and Basu (2000).

\subsection {Starless or Pre-protostellar?}

The question of whether to call these cores starless or pre-protostellar is a 
thorny one.  While these cores have no IRAS sources and so can all be called 
starless, the term pre-protostellar implies that the future evolution of these 
objects can be definitely predicted.  However, we can use the submillimeter 
observations of Ward-Thompson et al.\ (1994) to divide the sample into those 
cores detected by Ward-Thompson et al.\ at one or more wavelengths and those 
not detected at all (Table 6).  Of the 12 cores in the first group, 6 display 
blue asymmetry while the other six have no asymmetry.  The six cores that 
display blue asymmetry are the strongest in the submillimeter continuum.  Of 
the 5 cores Ward-Thompson et al.\ did not detect, 2 have symmetric profiles 
while we did not detect the other 3 in HCO$^{+}$.  The 2 detections of these 
cores were among the weakest lines observed.   We also include the NH$_{3}$ 
observations of Benson and Myers (1989) as a marker of dense core evolution.  
Of the 12 cores with submillimeter emission, all were observed in NH$_{3}$.  Of 
the 5 others, only one was observed.  Five of the cores in the first group have
since been mapped in the submillimeter continuum (Shirley et al. 2000).
The four most centrally condensed cores show blue asymmetry.  The least 
centrally condensed core, L1512, has no asymmetry in HCO$^{+}$.

There is a clear distinction between those cores with submillimeter 
continuum and those without. The first group has stronger line
emission that often shows what could be the beginning of infall.  
The group with relatively strong
submillimeter continuum, HCO$^{+}$, and NH$_{3}$ 
emission are the likeliest to be pre-protostellar in nature, consistent with 
the suggestion by Ward-Thompson et al.  (1994).

The cores with weak submillimeter continuum could eventually form stars
but were undetected simply because of lower column
density.  These cores may be in an even earlier evolution stage in
which cores are just beginning to form, or these cores may be forming
stars of lower mass.   
If these cores are weakly centrally concentrated,
they would not have a sufficient excitation temperature gradient for infall
asymmetry.  Further observations would be needed to resolve
the nature of these objects.  

\section {CONCLUSIONS}

We have observed 17 starless cores in HCO$^{+}$ $J=3-2$ in search of 
the spectral signature of pre-protostellar collapse.  These objects do seem to 
be a younger population than the Class 0 sources based on their narrow
line widths and weak peak temperatures.  We have observed blue 
asymmetric line profiles in 6 of these cores, and we suggest L1544, L1689B, B133 and L183 
as good protostellar collapse candidates and L1689A and L63 as worthy of 
further observations in the H$^{13}$CO$^{+}$ $J=3-2$ line.  
The blue excess of this sample is as prominent as in samples of 
``older" Class 0 and I sources, suggesting that infalling
 protostars are not exclusively to be found among the Class 0 sources.  
A population of likely pre-protostellar cores can be distinguished by
their strong submillimeter continuum and HCO$^{+}$ and NH$_{3}$ spectral
line emission.

However, recently Tafalla et al.\ (1998) have found that L1544 cannot be 
described as inside-out collapse.  If infall is happening in that core, it 
cannot be explained by any current infall model, suggesting that further study 
of these cores can tell us about the very beginning of the collapse process.

\acknowledgments

We would like to thank Tommy Greathouse, Wenbin Li, Byron Mattingly 
and Yancy Shirley for their help with observations.  We thank the referee
for helpful suggestions, one of which led us to reconsider the rest
frequency for the H$^{13}$CO$^+$ line. This work was supported by NSF
grant AST-9317567 and NASA grant NAG5-7203.

\clearpage
\begin{planotable}{lllllllll}
\tablewidth{6in}
\tablenum{1}
\tablecaption{List of Sources}
\tablehead{\colhead{Name} & \colhead{R.A.} & \colhead{Dec.} &
\colhead{Offpos\tablenotemark{a}} &\colhead{Distance} & \colhead{450 $\mu$m\tablenotemark{b}} & \colhead{800 $\mu$m} & \colhead{1.1 mm} & \colhead{1.3 mm}\\
& \colhead{(1950.0)} & \colhead{(1950.0)} & \colhead{(\arcsec)} & \colhead{(pc)}
& \colhead{(mJy)} & \colhead{(mJy)} & \colhead{(mJy)} & \colhead{(mJy)} }
\startdata
L1498 & 04:07:50.0 & 25:02:31 &  (-1200,0) & 140 & 700$\pm$80 & 120$\pm$18 & 35$\pm$6 & 10$\pm$2.5\nl
L1495D & 04:11:15.5 & 28:07:20 &(-900,0) & 140 & - & $<$135 & $<$44 & $<$30 \nl
L1506 & 04:15:30.3 & 25:13:22 & (-900,0) & 140 & - & $<$66 & $<$48 & -\nl
L1521A & 04:23:38.4 & 26:09:27 &(-900,0) & 140 & - & $<$110 & $<$80 & -\nl
L1517C & 04:51:35.9 & 30:30:00 & (-900,0)& 140 & 836$\pm$160 & 100$\pm$20 & $<$83 & $<$7.5\nl
L1517A & 04:51:54.8 & 30:28:53 &(-900,0) & 140 & 1280$\pm$330 & 105$\pm$18 & $<$60 & $<$5.4\nl
L1517D & 04:52:36.5 & 30:34:02 & (-900,0)& 140 & $<$4500 & $<$120 & $<$130 & -\nl
L1512 & 05:00:54.4 & 32:39:37 & (-900,0) & 140 & $<$6000 & 107$\pm$21 & 45$\pm$9 & $<$16\nl
L1544 & 05:01:13.1 & 25:06:36 & (-900,0) & 140 & 1300$\pm$240 & 450$\pm$58 & 193$\pm$30 & 46$\pm$4\nl
L1582A & 05:29:14.6 & 12:28:08 &(-900,0) & 140 & $<$1240 & 160$\pm$27 & $<$54 & $<$30\nl
L134A & 15:51:05.6 & -04:26:10 &(-900,0) & 150 & - & $<$60 & - & $<$163 \nl
L183 & 15:51:32.7 & -02:42:19 & (600,0) & 150 & $<$1500 & 269$\pm$30 & 108$\pm$26 & $<$134\nl
L1696A & 16:25:30.0 & -24:13:22 & (0,-1200) & 125 & 800$\pm$160 & 105$\pm$18 & 62$\pm$12 & $<$58\nl
L1689A & 16:29:10.5 & -24:57:22 & (0,-900) & 125 & 2200$\pm$300 & 290$\pm$45 & $<$102 & 54$\pm$15 \nl
L1689B & 16:31:47.0 & -24:31:45 & (0,-900) & 125 & $<$3000 & 362$\pm$40 & 140$\pm$34 & 134$\pm$11 \nl
L63 & 16:47:19.4 & -18:01:16 & (0,-900) & 125 & 1600$\pm$200 & 367$\pm$23 & $<$93 & $<$96 \nl 
B133 & 19:03:27.3 & -06:57:00 & (0,-600) & 400 & $<$1800 & 341$\pm$63 & $<$120 &$<$56 
\tablenotetext{a}{The off position used for position switching.}
\tablenotetext{b}{All photometry from Ward-Thompson et al.\ (1994).}
\end{planotable}

\clearpage
\begin{planotable}{llllll}
\tablewidth{6in}
\tablenum{2}
\tablecaption{List of Observed Lines}
\tablehead{\colhead{Molecule} & \colhead{Transition} & \colhead{Beamwidth} &
\colhead{$\eta_{mb}$} & \colhead{Resolution} & \colhead{Frequency} \\ & &
\colhead{(\arcsec)} & & \colhead{(\kms)} & \colhead{(MHz)}}
\startdata
H$^{13}$CO$^{+}$ & $J=3-2$ & 26 & 0.66 & 0.15 & 260255.617 (0.035) \nl
HCO$^{+}$ & $J=3-2$ & 26 & 0.66 & 0.16 & 267557.620 (0.01)
\end{planotable}

\clearpage
\begin{planotable}{lllll}
\tablewidth{6in}
\tablenum{3}
\tablecaption{Results}
\tablehead{\colhead{Source} & \colhead{Line} &
\colhead{$T_A^*$} & \colhead{$V_{LSR}$} & \colhead{$\Delta$V} \\ & &
\colhead{(K)} & \colhead{(\kms)} & \colhead{(\kms)}}
\startdata
L1498 & HCO$^{+}$ $J=3-2$ & 0.48$\pm$0.04 & 7.81$\pm$0.02 & 0.51$\pm$0.04 \nl
L1495D & HCO$^{+}$ $J=3-2$ & $<$0.05 & -- & -- \nl
L1506 & HCO$^{+}$ $J=3-2$ & 0.20$\pm$0.03 & 7.47$\pm$0.02 & 0.38$\pm$0.07 \nl
L1521A & HCO$^{+}$ $J=3-2$ & $<$0.05 & -- & -- \nl
L1517C & HCO$^{+}$ $J=3-2$ & 0.21$\pm$0.05 & 5.63$\pm$0.03 & 0.28$\pm$0.09 \nl
L1517A & HCO$^{+}$ $J=3-2$ & 0.26$\pm$0.05 & 5.72$\pm$0.04 & 0.70$\pm$0.10 \nl
L1517D & HCO$^{+}$ $J=3-2$ & $<$0.05 & -- & -- \nl
L1512 & HCO$^{+}$ $J=3-2$ & 0.41$\pm$0.04 & 7.05$\pm$0.02 & 0.45$\pm$0.04 \nl
L1544 & H$^{13}$CO$^{+}$ $J=3-2$ & 0.09$\pm$0.02 & 7.18$\pm$0.04 & 0.50$\pm$0.10 \nl
- & HCO$^{+}$ $J=3-2$ & 1.11$\pm$0.07 & 6.96$\pm$0.06 & 0.60$\pm$0.12 \nl
- & - & 0.91$\pm$0.07 & 7.44$\pm$0.06 & - \nl
L1582A & HCO$^{+}$ $J=3-2$ & 0.77$\pm$0.05 & 10.06$\pm$0.02 & 0.83$\pm$0.04 \nl
L134A & HCO$^{+}$ $J=3-2$ & 0.23$\pm$0.04 & 2.93$\pm$0.04 & 0.44$\pm$0.11 \nl
L183 & H$^{13}$CO$^{+}$ $J=3-2$ & 0.08$\pm$0.02 & 2.67$\pm$0.04 & 0.43$\pm$0.09 \nl
- & HCO$^{+}$ $J=3-2$ & 0.28$\pm$0.01 & 2.27$\pm$0.08 & 0.96$\pm$0.16 \nl
- & -- & 0.25$\pm$0.01 & 2.75$\pm$0.08 & -- \nl
L1696A & HCO$^{+}$ $J=3-2$ & 0.91$\pm$0.06 & 3.45$\pm$0.01 & 0.43$\pm$0.03 \nl
L1689A & H$^{13}$CO$^{+}$ $J=3-2$ & $<$0.03 & -- & -- \nl
-- & HCO$^{+}$ $J=3-2$ & 0.53$\pm$0.03 & 3.57$\pm$0.02 & 1.13$\pm$0.04 \nl
L1689B & H$^{13}$CO$^{+}$ $J=3-2$ & 0.23$\pm$0.03 & 3.41$\pm$0.02 & 0.50$\pm$0.06 \nl
-- & HCO$^{+}$ $J=3-2$ & 1.12$\pm$0.07 & 3.33$\pm$0.01 & 0.58$\pm$0.04 \nl
L63 & H$^{13}$CO$^{+}$ $J=3-2$ & 0.12$\pm$0.03 & 5.73$\pm$0.06 & 0.71$\pm$0.24 \nl 
-- & HCO$^{+}$ $J=3-2$ & 1.07$\pm$0.03 & 5.53$\pm$0.08 & 0.60$\pm$0.15 \nl
B133 & H$^{13}$CO$^{+}$ $J=3-2$ & 0.12$\pm$0.03 & 12.27$\pm$0.04 & 0.38$\pm$0.07 \nl
-- & HCO$^{+}$ $J=3-2$ & 0.63$\pm$0.03 & 12.01$\pm$0.01 & 0.57$\pm$0.03
\end{planotable}

\clearpage
\begin{planotable}{lll}
\tablewidth{6in}
\tablenum{4}
\tablecaption{Line Asymmetry}
\tablehead{\colhead{Source} & \colhead{Blue/Red} & 
\colhead{Asymmetry}}
\startdata
L1498 & -- &  -0.04$\pm$0.09 \nl
L1512 & -- &  -0.32$\pm$0.12 \nl
L1544 & 1.22$\pm$0.12 & -0.44$\pm$0.17 \nl
L1582A & -- &  -0.35$\pm$0.06 \nl
L134A & -- &  (0.95$\pm$0.21) \nl
L183 & 1.12$\pm$0.06 &  -0.93$\pm$0.28 \nl
L1696A & -- &  0.12$\pm$0.06 \nl
L1689A & 1.23$\pm$0.11 & -- \nl
L1689B & 2.48$\pm$0.38 & -0.16$\pm$0.05 \nl
L63 & 1.55$\pm$0.08 & -0.28$\pm$0.16 \nl
B133 & 3.14$\pm$0.45 & -0.68$\pm$0.17 
\tablenotetext{a} {The asymmetry for L134A was calculated from
N$_{2}$H$^{+}$ spectra (Benson et al.\ 1999) observed 115\arcsec\
 from our HCO$^{+}$ position.} 
\end{planotable}

\clearpage
\begin{planotable}{lllll}
\tablewidth{6in}
\tablenum{5}
\tablecaption{Asymmetry Statistics}
\tablehead{\colhead{Method} & \colhead{Blue} & \colhead{Symmetric} &
\colhead{Red} & \colhead{Excess}}
\startdata
Asymmetry & 6 & 3 & 0 & 0.67 \nl
Profiles & 6 & 8 & 0 & 0.43
\end{planotable}

\clearpage
\begin{planotable}{lllll}
\tablewidth{6in}
\tablenum{6}
\tablecaption{Criteria for Pre-Protostellar Cores}
\tablehead{\colhead{Source} & \colhead{Continuum} & \colhead{HCO$^{+}$} 
& \colhead{NH$_{3}$} & \colhead{Blue profile}}
\startdata
L1498 & Y & Y & Y & N \nl
L1517C & Y & Y & Y & N \nl
L1517A & Y & Y & Y & N \nl
L1512 & Y & Y & Y & N \nl
L1544 & Y & Y & Y & Y \nl
L1582A & Y & Y & Y & N \nl
L183 & Y & Y & Y & Y \nl
L1696A & Y & Y & Y & N \nl
L1689A & Y & Y & Y & Y \nl
L1689B & Y & Y & Y & Y \nl
L63 & Y & Y & Y & Y \nl
B133 & Y & Y & Y & Y \nl
\nl
L1495D & N & N & N & N \nl
L1506 & N & Y & N & N \nl
L1521A & N & N & N & N \nl
L1517D & N & N & N & N \nl
L134A & N & Y & Y & N
\end{planotable}

\clearpage
\begin{figure}[htb!] 
\centering 
\epsfxsize=4.5in 
\epsfysize=6in 
\epsfbox{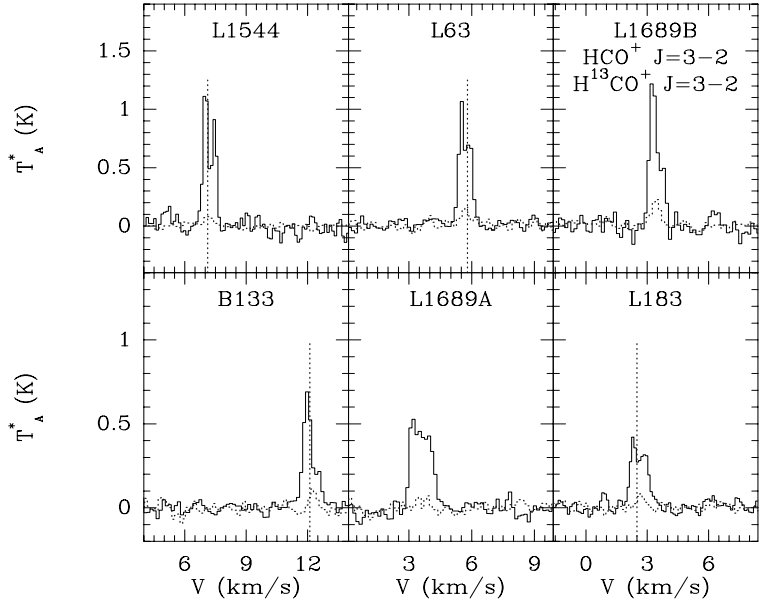} 
\caption{HCO$^{+}$ and H$^{13}$CO$^{+}$ $J=3-2$ spectra toward the center of 
six starless cores.  The HCO$^{+}$ $J=3-2$ spectra is the solid line 
and the H$^{13}$CO$^{+}$ $J=3-2$ spectra is the dashed line.  The dashed 
vertical line is the velocity of the N$_{2}$H$^{+}$ line of Benson et al.\ 
(1998).}
\end{figure}

\begin{figure}[htb!] 
\centering 
\epsfxsize=4.5in 
\epsfysize=6in 
\epsfbox{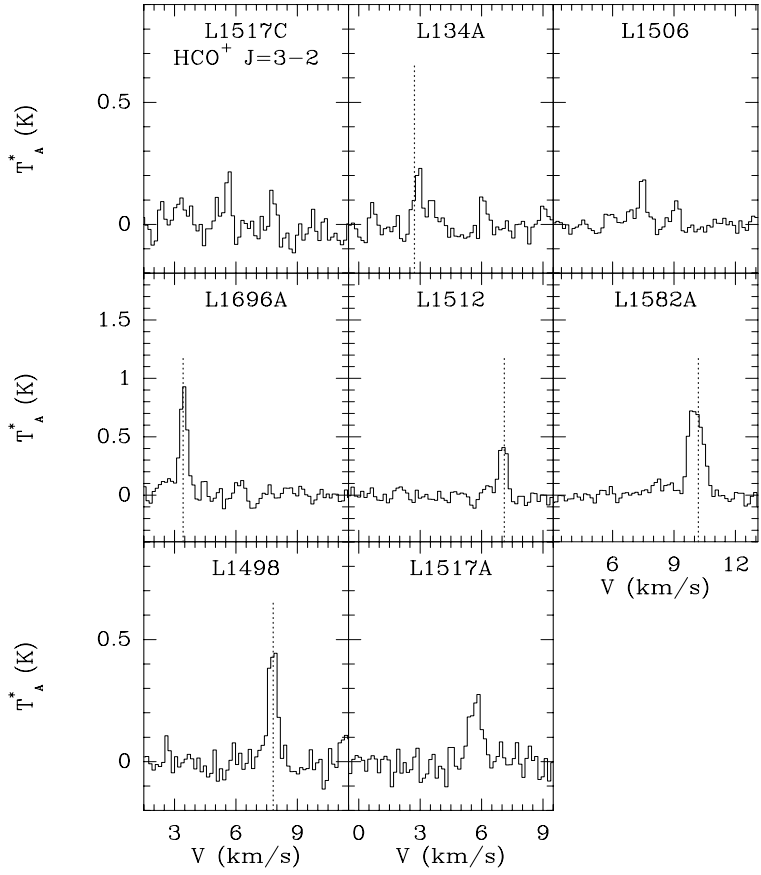} 
\caption{HCO$^{+}$ $J=3-2$ spectra toward the center of eight starless cores.  
The dashed vertical line is the velocity of the N$_{2}$H$^{+}$ line of
Benson et al.\ (1998).}
\end{figure}

\begin{figure}[htb!] 
\centering 
\epsfxsize=4.5in 
\epsfysize=6in 
\epsfbox{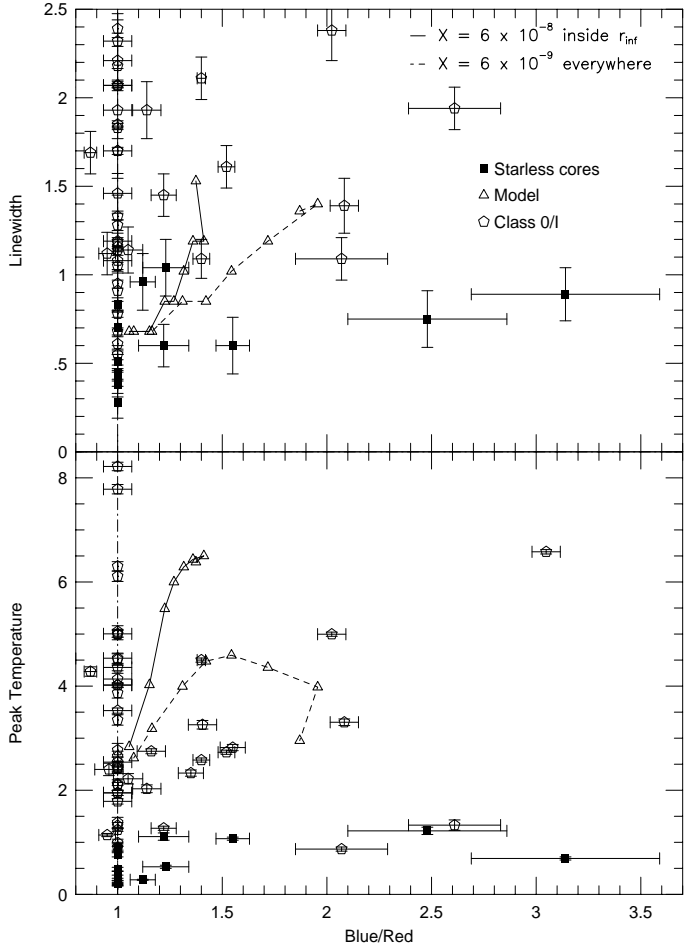}
\caption{The top plot is the peak temperature of the HCO$^{+}$ $J=3-2$ line
versus blue-red asymmetry and the bottom plot is linewidth of the 
HCO$^{+}$ $J=3-2$ line versus blue-red asymmetry.  Open triangles are
points from the evolutionary models of Gregersen et al.\ (1997).  Open trianglesconnected by a solid line are from infall models with X(HCO$^{+}$) = 6 $\times$
10$^{-8}$ inside the infall radius; those connected by a dashed line have
X(HCO$^{+}$) = 6 $\times$ 10$^{-9}$ throughout.  Filled
squares are the starless cores and the open pentagons are previously
observed Class 0 and Class I sources (Gregersen et al.\ 1997, 2000 
respectively).}
\end{figure}
 
\begin{figure}[htb!] 
\centering 
\epsfxsize=4.5in 
\epsfysize=6in
\epsfbox{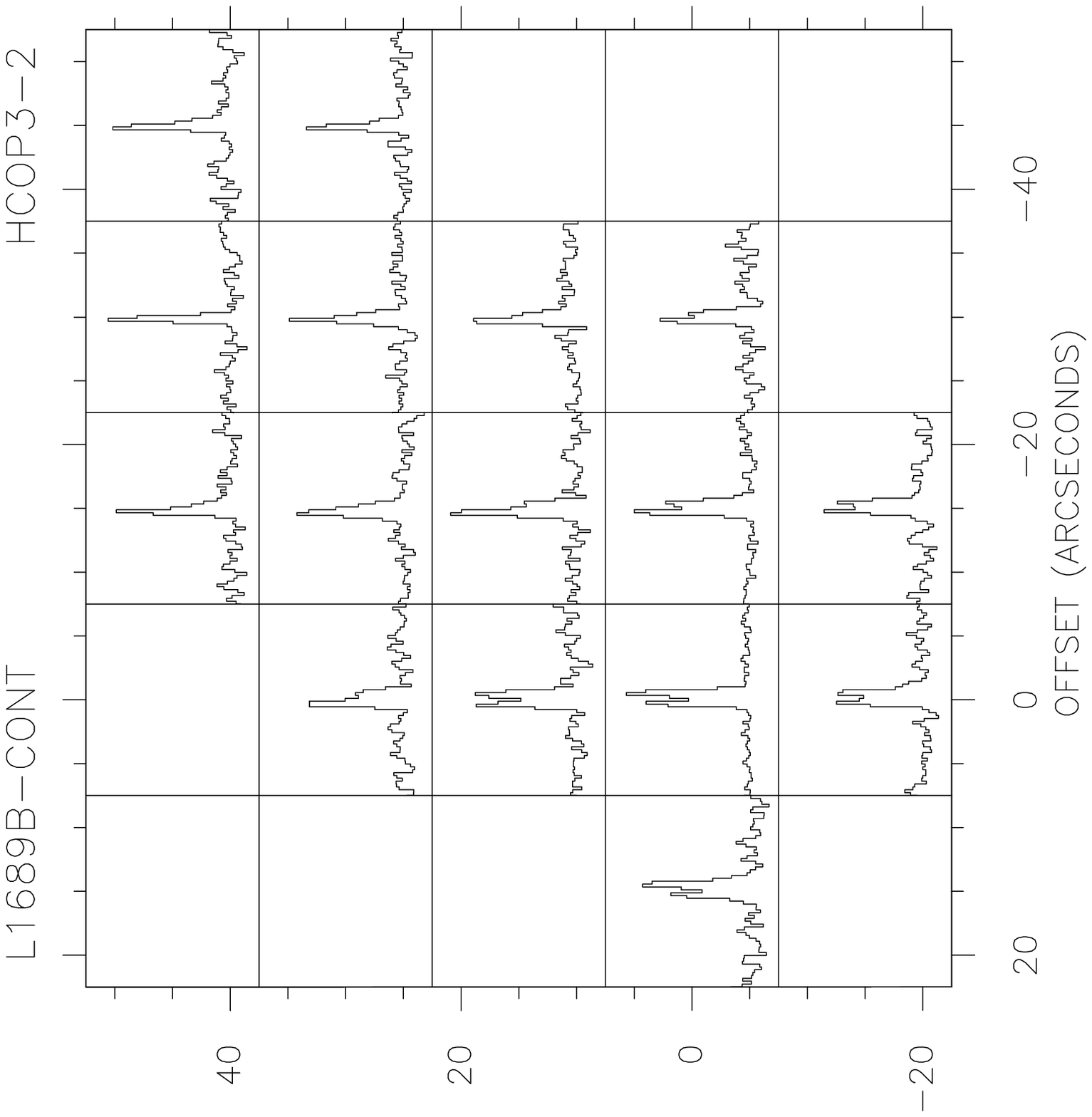} 
\caption{Map of HCO$^{+}$ $J=3-2$ spectra in L1689B.  The velocity scale is 
from $-$1.5 to 8.5 \kms\ and the temperature scale is from $-$0.3 to 1.4 K.}
\end{figure}  
  
\begin{figure}[htb!] 
\centering 
\epsfxsize=4.5in 
\epsfysize=6in 
\epsfbox{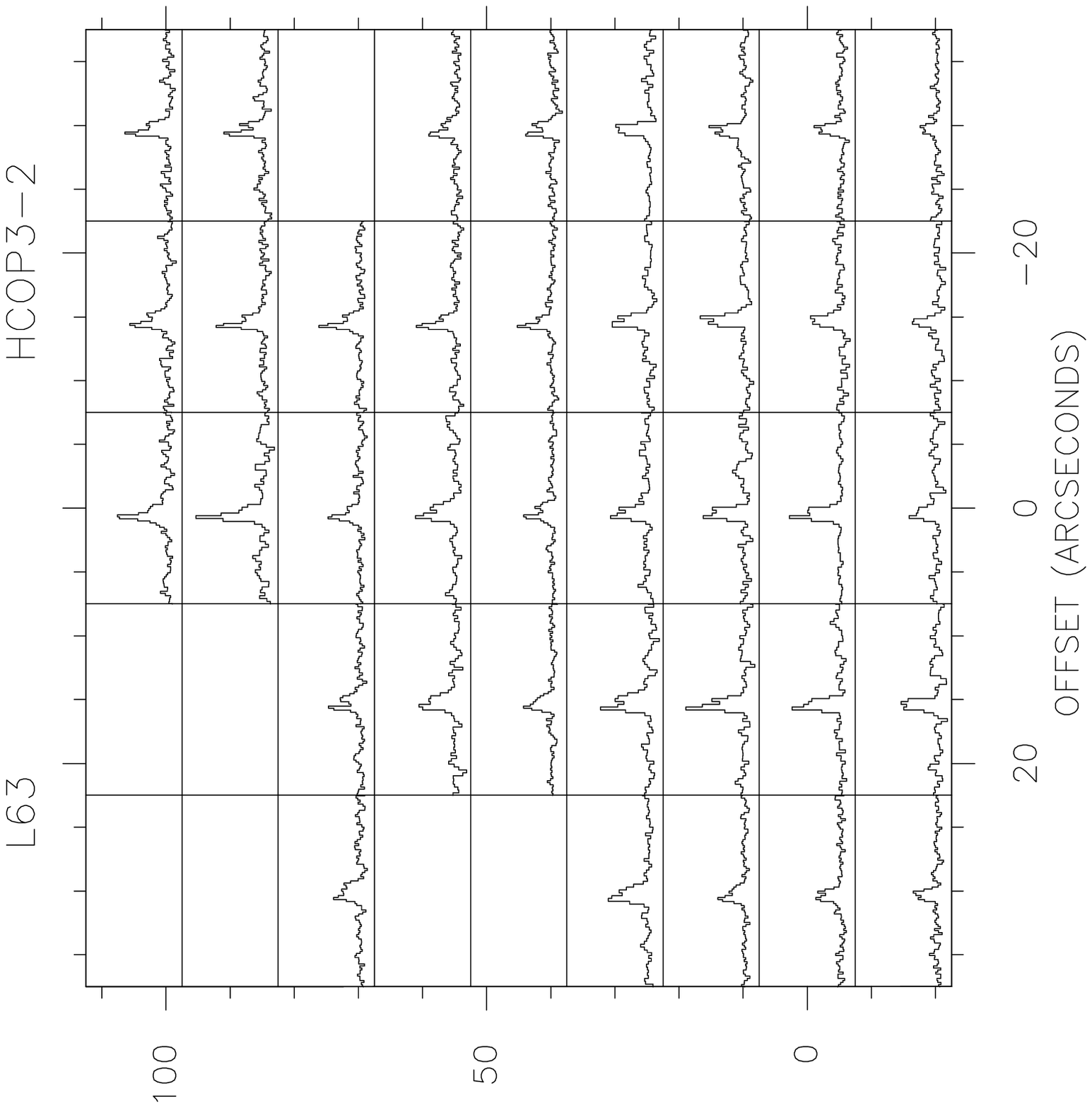} 
\caption{Map of HCO$^{+}$ $J=3-2$ spectra in L63.  The velocity scale is  
from 1 to 11 \kms\ and the temperature scale is from $-$0.3 to 1.7 K.}
\end{figure}

\begin{figure}
\centering 
\epsfxsize=4.5in 
\epsfysize=6in 
\epsfbox{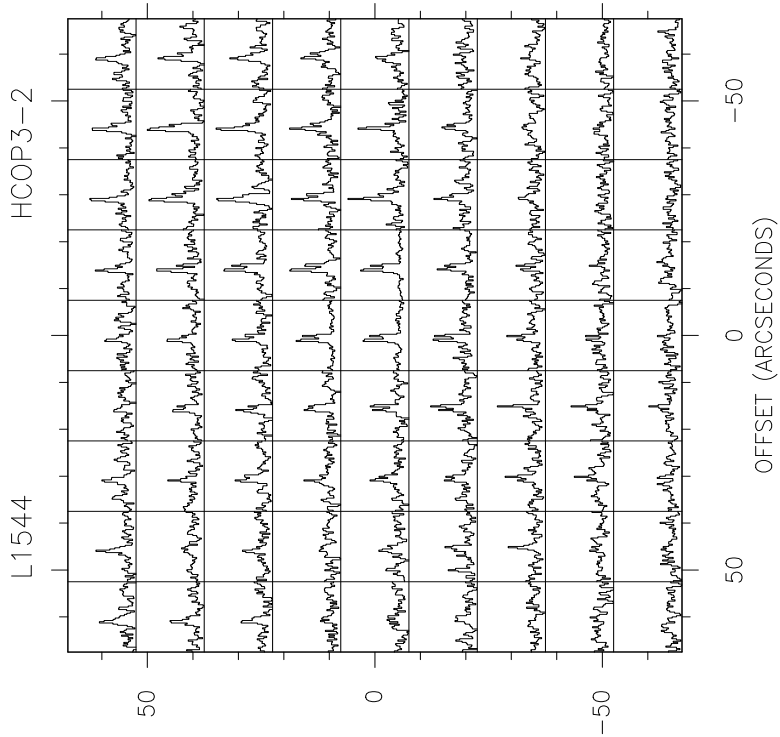}
\caption{Map of HCO$^{+}$ $J=3-2$ spectra in L1544.  The velocity scale is
from 2.5 to 12.5 \kms\ and the temperature scale is from $-$0.3 to 1.8 K.}
\end{figure}

\end{document}